\documentstyle[12pt,epsfig]{article}
\begin{document}
\def \beq {\begin{equation}}
\def \eeq {\end{equation}}
\def \eeqn {\end{equation}\noindent}
\def \bes {\begin{eqnarray}}
\def \ees {\end{eqnarray}}
\def \eesn {\end{eqnarray}\noindent}
\def \nn {\nonumber}
\def\mum{\,$\mu$m}

\large

\begin{center}
{\Large {\bf Higher order conductivity corrections
\\ 
to the Casimir force }}
\vskip 6mm

{\large
V.B.~Bezerra,  
G.L.~Klimchitskaya\footnote{On leave from: North-West Polytechnical
Institute, St.~Petersburg, Russia},
V.M.~Mostepanenko\footnote{On leave from: A.~Friedmann
Laboratory for Theoretical Physics, St.Petersburg, 
Russia}}\\[5mm]

{\normalsize
{\it Departamento de F\'{\i}sica, Universidade Federal da Para\'{\i}ba, \\
Caixa Postal 5008, CEP 58059-970
Jo\~{a}o Pessoa, Pb-Brazil}}

\end{center}
\vskip 6mm
\begin{abstract}
 The finite conductivity corrections to the Casimir force 
in two configurations are calculated in 
the third and fourth orders in relative penetration depth
of electromagnetic zero oscillations into the metal.
The obtained analytical perturbation results are
compared with recent computations. Applications to
the modern experiments are discussed.
\end{abstract}
\vskip 8mm
PACS codes: 03.70.+k, 12.20.Ds, 78.20.-e \hfill \\
Keywords: Casimir effect, finite conductivity
corrections, plasma model 
\\[10mm]
Corresponding author: V.M. Mostepanenko, Departamento de F\'{\i}sica, \\UFPB, Caixa Postal 5008, CEP~58.059-970 
Jo\~{a}o Pessoa, Pb-Brazil\\
Tel: 55(83)216-7529 \hfill \\
Fax: 55(83)216-7422 \hfill \\
E-mail: mostep@fisica.ufpb.br \hfill
\newpage
\section{Introduction}

\hspace*{\parindent}
In 1998 just 50 years have passed after the publication
of the famous paper by Casimir [1] (see the Proceedings
[2] especially devoted to this event). Accidentally the
experimental interest in the Casimir effect was
rekindled at the same time. In Ref.[3] the torsion
pendulum was used to measure the Casimir force between
$Cu$ plus $Au$ coated quartz plate, and a spherical lens
in a distance range from 0.6\mum\  to 6\mum. The accuracy
of order 5\% was claimed in [3] for the agreement of the
measurement with theory. In Refs.[4--6] the Casimir
force between $Al$ plus $Au/Pd$ coated disk and a sphere
was measured for surface separations between 0.1\mum\ 
to 0.9\mum\ using the Atomic Force Microscope. The
deviation between theory and experiment was shown to be
of around 1\% at the smallest surface separation [4,6,7].
The obtained experimental results and the extent of their
agreement with theory were used to establish stronger
constraints for the parameters of hypothetical long-range
interactions predicted by the unified gauge theories,
supersymmetry and supergravity [8--11].

To be confident that  data fit theory at a level of about
several percent, the different corrections to the ideal
expression for the Casimir force should be taken into
account. The main contribution is given by the corrections
due to finite conductivity of the boundary metal, its
roughness and due to non-zero temperature (see [12] for
review). Experimental data of [3] do not support the
presence of any of these corrections although they should
contribute at a level of 5\%. By contrast, in [4,6] the
surface roughness and finite conductivity corrections are
of great concern (the temperature corrections are
negligible in the measurement range of [4,6]).

The subject of the present paper is the calculation of
higher order finite conductivity corrections to the
Casimir force in relative penetration depth of
electromagnetic zero oscillations into the metal. We
consider configurations of two plane parallel plates and
a sphere above a plate. The first order finite conductivity
correction was found in [13] for configuration of two plane
parallel plates with an error in numerical coefficient
corrected in [14]. Later the correct result was reobtained
in [15]. Second order correction was firstly found in [16]
(see also [12]). It was modified for the configuration of
a sphere above a disk in [17] by the use of Proximity
Force Theorem (PFT) [18]. The results of [16,17] for 
the Casimir
force up to the second power in relative penetration
depth are in common use when discussing the recent
experiments (see, e.g., [4,6,7,19--21]). In [7,10] the
third and the fourth order corrections were obtained
approximately from the interpolation formula. They
allowed to achieve the excellent agreement between theory
and experiment.

In [19] numerical calculation of the Casimir force with
account of finite conductivity has been attempted based
on the tabulated data for the comp\-lex dielectric
permittivity as a function of frequency. The same
computation was repeated in [21] with the diverged
results. The reason of these dif\-fe\-ren\-ces 
was interpreted
in [21] as the invalid manipulation of optical data in
[19]. Our analytical calculation of higher order 
conductivity corrections agrees with the results of [21] 
in the application range of perturbation approach. As
shown below the perturbation results obtained in the
context of plasma model are valid with rather high
accuracy when the distance between the test bodies is
larger than the plasma wavelength (not much larger as
advocated in [20]). 
This gives the
possibility, in some instance, to use the plasma model
for the distances of order or even less than the
characteristic absorption wavelength of test body material.

The paper is organized as follows. In Sec.~2 the general
finite results for the Casimir energy density and force
are briefly presented. Sec.~3 contains derivation of the
third and fourth order conductivity corrections. 
In Sec.~4 the
obtained perturbation results are compared with numerical
calculations. Sec.~5 contains conclusions and discussion.

\section{Casimir energy density and force between
realistic materials}

\hspace*{\parindent}
Let us consider two semi-infinite solids with dielectric
permittivity $\varepsilon(\omega)$ separated by a 
plane-parallel gap of width $a$. The surfaces of the bodies
are planes $z=0, a$. The Casimir energy density
and force acting between these bodies can be found most
simply following [22,23] (see also [24] for the
multilayered walls). Some additional clarification is
given here in the case in which the finite energy
density rather than force is the subject of interest.

The mode frequencies of electromagnetic field are found
from Maxwell equations supplemented by the standard
boundary conditions at $z=0, a$. Two types of such
frequencies 
$\omega_{\mbox{\footnotesize{\boldmath$k$}},n}^{(1,2)}$
(surface modes [25]) corresponding to two polarizations 
of the electric field are the solutions of the equations
\bes
&&
\Delta{\!}^{(1)}(
\omega_{\mbox{\footnotesize{\boldmath$k$}},n}^{(1)})
\equiv\varepsilon^2 K_{\varepsilon}^2
e^{-K_{\varepsilon}a}\left[
\left(K_{\varepsilon}+\varepsilon K\right)^2
e^{Ka}-
\left(K_{\varepsilon}-\varepsilon K\right)^2
e^{-Ka}\right]=0,\nn\\
&&
\Delta{\!}^{(2)}(
\omega_{\mbox{\footnotesize{\boldmath$k$}},n}^{(2)})
\equiv K_{\varepsilon}^2
e^{-K_{\varepsilon}a}\left[
\left(K_{\varepsilon}+ K\right)^2 e^{Ka}-
\left(K_{\varepsilon}- K\right)^2 e^{-Ka}
\right]=0.
\label{1}
\eesn
Here the following notations are introduced
\beq
K^2=k^2-\frac{\omega^2}{c^2}, \qquad
K_{\varepsilon}^2=k^2-\varepsilon\frac{\omega^2}{c^2},
\label{2}
\eeqn
and {\boldmath$k$} is the two-dimensional 
propagation vector
in the $xy$-plane.

The infinite zero-point energy of the electromagnetic field
between the plates is given by [23,24]
\beq
E(a)=\frac{1}{2}\hbar
\sum\limits_{\mbox{\footnotesize{\boldmath$k$}},n}
\left(\omega_{\mbox{\footnotesize{\boldmath$k$}},n}^{(1)}
+\omega_{\mbox{\footnotesize{\boldmath$k$}},n}^{(2)}
\right),
\label{3}
\eeqn
where the sum in continuous index {\boldmath$k$} is
actually an integral. Introducing the length $L$ for the
$x, y$ sides of the plates we obtain the vacuum energy
density
\beq
{\cal{E}}(a)=\frac{E(a)}{L^2}=
\frac{\hbar}{4\pi}
\int\limits_{0}^{\infty}\! k\,dk
\sum\limits_{n}
\left(\omega_{\mbox{\footnotesize{\boldmath$k$}},n}^{(1)}
+\omega_{\mbox{\footnotesize{\boldmath$k$}},n}^{(2)}
\right),
\label{4}
\eeqn
which is also infinite.

Summation in (\ref{4}) over the solutions of (\ref{1})
can be performed with the help of the argument principle
\beq
\sum\limits_{n}
\omega_{\mbox{\footnotesize{\boldmath$k$}},n}^{(1,2)}
=\frac{1}{2\pi i}\left[
\int\limits_{i\infty}^{-i\infty}\!\!
\omega d\ln\Delta{\!}^{(1,2)}(\omega)+
\int\limits_{C{\!}_{+}}\!
\omega d\ln\Delta{\!}^{(1,2)}(\omega)\right],
\label{5}
\eeqn
where contour $C{\!}_{+}$ is a right semicircle of
infinite radius in complex $\omega$-plane with a center 
at the origin (note that the functions 
$\Delta{\!}^{(1,2)}(\omega)$ have no poles).

The integral over $C{\!}_{+}$ can be simply calculated when
it is considered that 
\beq
\lim\limits_{\omega\to\infty}\varepsilon(\omega)=1,
\qquad
\lim\limits_{\omega\to\infty}
\frac{d\varepsilon(\omega)}{d\omega}=0
\label{6}
\eeqn
along the arbitrary radial direction in complex plane.
The result (infinite) does not depend on $a$. It is
given by
\beq
\int\limits_{C{\!}_{+}}\!
\omega d\ln\Delta{\!}^{(1,2)}(\omega)=
3\int\limits_{C{\!}_{+}}\!
d\omega.
\label{7}
\eeq

Introducing a new variable $\xi=-i\omega$ in (\ref{5}),
(\ref{7}) and performing a partial integration one
obtains
\beq
\sum\limits_{n}
\omega_{\mbox{\footnotesize{\boldmath$k$}},n}^{(1,2)}
=\frac{1}{2\pi}
\int\limits_{-\infty}^{\infty}\!\!
\xi d\ln\Delta{\!}^{(1,2)}(i\xi)+
\frac{3}{2\pi}
\int\limits_{C{\!}_{+}}\!
d\xi.
\label{8}
\eeq

Now let us turn to the removing of divergencies (this
important point was not discussed in [22--24]). It is
apparent that for the infinitely remote plates the
regularized physical vacuum energy density should
vanish [12]. In the limit $a\to\infty$ we have from
(\ref{8})
\beq
\left(\sum\limits_{n}
\omega_{\mbox{\footnotesize{\boldmath$k$}},n}^{(1,2)}
\right)_{\infty}
=\frac{1}{2\pi}
\int\limits_{-\infty}^{\infty}\!\!
\xi d\ln\Delta{\!}_{\ \infty}^{(1,2)}(i\xi)+
\frac{3}{2\pi}
\int\limits_{C{\!}_{+}}\!
d\xi.
\label{9}
\eeqn
where by the use of (\ref{1}), it follows
\beq
\Delta{\!}_{\ \infty}^{(1)}=
\varepsilon^2 K_{\varepsilon}^2
e^{-K_{\varepsilon}a}
\left(K_{\varepsilon}+\varepsilon K\right)^2
e^{Ka},
\quad
\Delta{\!}_{\ \infty}^{(2)}=
 K_{\varepsilon}^2 e^{-K_{\varepsilon}a}
\left(K_{\varepsilon}+ K\right)^2 e^{Ka}.
\label{10}
\eeq

For a regularized quantity the result is
\beq
\left(\sum\limits_{n}
\omega_{\mbox{\footnotesize{\boldmath$k$}},n}^{(1,2)}
\right)_{reg}\equiv
\sum\limits_{n}
\omega_{\mbox{\footnotesize{\boldmath$k$}},n}^{(1,2)}
-\left(\sum\limits_{n}
\omega_{\mbox{\footnotesize{\boldmath$k$}},n}^{(1,2)}
\right)_{\infty}=
\frac{1}{2\pi}
\int\limits_{-\infty}^{\infty}\!\!
d\xi \ln\frac{\Delta{\!}^{(1,2)}
(i\xi)}{\Delta{\!}_{\ \infty}^{(1,2)}(i\xi)},
\label{11}
\eeqn
where we are guided by the argument of infinitely remote
plates. 

Substituting the regularized quantities (\ref{11}) into
(\ref{4}) instead of (\ref{8}) we obtain the final
expression for the Casimir energy density between plates
\beq
{\cal{E}}_{reg}(a)=\frac{\hbar}{4\pi^2}
\int\limits_{0}^{\infty}\! kdk
\int\limits_{0}^{\infty}\! d\xi\left[
\ln\tilde\Delta{\!}^{(1)}(i\xi)+
\ln\tilde\Delta{\!}^{(2)}(i\xi)\right],
\label{12}
\eeqn
where
\beq
\tilde\Delta{\!}^{(1)}\equiv
\frac{\Delta{\!}^{(1)}}{\Delta{\!}_{\ \infty}^{(1)}}=1-
\frac{(K_{\varepsilon}-\varepsilon K)^2}{(K_{\varepsilon}
+\varepsilon K)^2}e^{-2Ka},
\quad
\tilde\Delta{\!}^{(2)}\equiv
\frac{\Delta{\!}^{(2)}}{\Delta{\!}_{\ \infty}^{(2)}}=1-
\frac{(K_{\varepsilon}- K)^2}{(K_{\varepsilon}
+ K)^2}e^{-2Ka},
\label{13}
\eeqn
and also use was made of the fact
that $\tilde\Delta{\!}^{(1,2)}$
are even functions of $\xi$.

Notice that in [23] no finite expression for the energy
density was obtained. In [24] the omission of infinities
was performed implicitly without a physical
 justification. To illustrate this, in [24] instead of
Eqs.(\ref{1}) the result of their division by the terms
containing $\exp(Ka)$ was used in spite of the fact that
on $C_{\! +}$ such operation is the division by infinity.
Fortunately, this operation did not influence the final
result for the energy density obtained in [24] which is
perfectly correct.

One can obtain the Casimir force between plates from
(\ref{12})
\bes
&&
F_p(a)=-\frac{\partial {\cal{E}}_{reg}(a)}{\partial a}=
-\frac{\hbar}{2\pi^2}
\int\limits_{0}^{\infty}\! kdk
\int\limits_{0}^{\infty}\! d\xi K\left\{\left[
\frac{(K_{\varepsilon}+\varepsilon K)^2}{(K_{\varepsilon}
-\varepsilon K)^2}e^{2Ka}-1\right]^{-1}\right.
\nn\\
&&\phantom{aaaaaaaaaaaaaaaaaaaaaa}
+\left.\left[
\frac{(K_{\varepsilon}+ K)^2}{(K_{\varepsilon}
- K)^2}e^{2Ka}-1\right]^{-1}\right\},
\label{14}
\eesn
which is exactly equivalent to Lifshitz result [23,26].
Using PFT [18] we obtain from 
(\ref{12}) the Casimir force acting between a plane plate
and a spherical lens or a sphere which is given by
\beq
F_l(a)=2\pi R{\cal{E}}_{reg}(a)=\frac{\hbar R}{2\pi}
\int\limits_{0}^{\infty}\! kdk
\int\limits_{0}^{\infty}\! d\xi\left[
\ln\tilde\Delta{\!}^{(1)}(i\xi)+
\ln\tilde\Delta{\!}^{(2)}(i\xi)\right].
\label{15}
\eeq

Both Eqs.(\ref{14}) and (\ref{15}) are used below to
calculate higher order conductivity corrections to the
Casimir force between realistic metals in two
configurations under consideration.

\section{Higher order conductivity corrections}

\hspace*{\parindent}
It is common knowledge that the dominant contribution to
the Casimir force comes from frequencies 
$\xi\sim c/a$. We consider the micrometre domain with
$a$ from a few tenths of a micrometre to around a
hundred micrometers. Here the dominant frequencies are
of visible light and infrared optics. In this domain, the
plasma model works well and the dielectric permittivity
of a metal can be presented as
\beq
\varepsilon(\omega)=1-\frac{\omega_p^2}{\omega^2},
\qquad
\varepsilon(i\xi)=1+\frac{\omega_p^2}{\xi^2},
\label{16}
\eeqn
where the plasma frequency $\omega_p$ is different for
different metals.

The case of plane parallel plates will be our initial
concern. Introducing new variables $p$ and $x$
according to
\beq
k^2=\frac{\xi^2}{c^2}(p^2-1),\qquad
\xi=\frac{cx}{2pa}
\label{17}
\eeqn
we transform Eq.(\ref{14}) into the form
\beq
F_p(a)=
-\frac{\hbar c}{32\pi^2 a^4}
\int\limits_{0}^{\infty}\! x^3dx
\int\limits_{1}^{\infty}\! \frac{dp}{p^2}
\left\{\left[
\frac{(s+p\varepsilon)^2}{(s-p\varepsilon )^2}
e^{x}-1\right]^{-1}+\left[
\frac{(s+ p)^2}{(s-p)^2}e^{x}-1\right]^{-1}\right\},
\label{18}
\eeqn
where 
\beq
s\equiv\sqrt{\varepsilon -1+p^2}.
\label{19}
\eeq

Let us expand the expression under the integral with 
respect
to $p$ in powers of a small parameter
\beq
\alpha\equiv\frac{\xi}{\omega_p}=
\frac{c}{2\omega_p a}\cdot \frac{x}{p}=
\frac{\delta_0}{a}\cdot\frac{x}{2p},
\label{20}
\eeqn
where $\delta_0=\lambda_p/(2\pi)$ is the effective
penetration depth of the electromagnetic oscillations into
the metal. Note that in terms of this parameter
$\varepsilon(\omega)= 1+ (1/\alpha^2)$.

After the straightforward calculations one obtains
\bes
&&
\left[
\frac{(s+p\varepsilon)^2}{(s-p\varepsilon )^2}
e^{x}-1\right]^{-1}=
\frac{1}{e^x-1}\left[\vphantom{\frac{A}{p^4}}
1-\frac{4A}{p}\alpha +
\frac{8A}{p^2}(2A-1)\alpha^2\right.
\nn\\
&&\phantom{aaaaaaaa}
+\frac{2A}{p^3}(-6+32A-32A^2+2p^2-p^4)\alpha^3
\label{21}\\
&&\phantom{aaaaaaaa}\left.
+\frac{8A}{p^4}(2A-1)(2-16A+16A^2-2p^2+p^4)\alpha^4
+O(\alpha^5)\right],
\nn
\eesn 
where $A\equiv e^x/(e^x-1)$.

In perfect analogy, the other contribution from (\ref{18})
is 
\bes
&&
\left[
\frac{(s+p)^2}{(s-p)^2}
e^{x}-1\right]^{-1}=
\frac{1}{e^x-1}\left[
1-4Ap\alpha +
8A(2A-1)p^2\alpha^2\right.
\nn\\
&&\phantom{aaaaaaaa}
+2A(-5+32A-32A^2)p^3\alpha^3
\label{22}\\
&&\phantom{aaaaaaaa}\left.
+8A(1+18A-48A^2+32A^3)p^4\alpha^4
+O(\alpha^5)\right]
\nn
\eesn 
(note that this expression actually does not depend on
$p$ due to (\ref{20})).

After substitution of (\ref{21}), (\ref{22}) into
(\ref{18}) all integrals with respest to $p$ have the
form $\int_{0}^{\infty}dpp^{-k}$ with $k\geq 2$ and are
calculated immediately. The integrals with respect to
$x$ have the form
\beq
\int\limits_{0}^{\infty}dx
\frac{x^ne^{mx}}{(e^x-1)^{m+1}}
\label{23}
\eeqn
and can be easily calculated with the help of [27].
Substituting their values into (\ref{18}) we obtain
after some transformations the Casimir force between
metallic plates with finite conductivity corrections
up to the fourth power in relative penetration depth
\bes
&&
F_p(a)=F_p^{(0)}(a)\left[
1-\frac{16}{3}\frac{\delta_0}{a}+
24\frac{\delta_0^2}{a^2}-
\frac{640}{7}\left(1-\frac{\pi^2}{210}\right)
\frac{\delta_0^3}{a^3}\right.
\nn\\
&&
\phantom{aaaaaaaaaaaaa}\left.+
\frac{2800}{9}\left(1-\frac{163\pi^2}{7350}\right)
\frac{\delta_0^4}{a^4}\right],
\label{24}
\eesn
where $F_p^{(0)}(a)\equiv -(\pi^2\hbar c)/(240a^4)$.

As was mentioned in the Introduction, the first order
correction in (\ref{24}) was obtained in [13--15].
 The second order correction was
obtained in [16] (see also [12]). The third and fourth
order corrections which are obtained here are important
for the recent Casimir force measurements (see Sec.4).

Now let us turn to the configuration of a lens or a sphere
above a plate. Introducing the new variable (\ref{17})
into (\ref{15}) we get the Casimir force 
\beq
F_l(a)=
\frac{\hbar cR}{16\pi a^3}
\int\limits_{0}^{\infty}\! x^2dx
\int\limits_{1}^{\infty}\! \frac{dp}{p^2}
\left\{\ln\left[1-
\frac{(s-p\varepsilon)^2}{(s+p\varepsilon )^2}
e^{-x}\right]+\ln\left[1-
\frac{(s-p)^2}{(s+p)^2}e^{-x}\right]\right\}.
\label{25}
\eeq

Bearing in mind the further expansions it is convenient
to perform in (\ref{25}) integration by parts with respect
to $x$. The result is
\bes
&&
F_l(a)=
-\frac{\hbar cR}{48\pi a^3}
\int\limits_{0}^{\infty}\! x^3dx
\int\limits_{1}^{\infty}\! \frac{dp}{p^2}
\left[
\frac{(s-p\varepsilon)^2-(s+p\varepsilon )^2
\frac{\partial}{\partial x}\frac{(s-
p\varepsilon)^2}{(s+p\varepsilon )^2}}{(s+
p\varepsilon)^2 e^x-(s-p\varepsilon )^2}
\right.
\nn\\
&&\phantom{aaaaaaaaaaaaaaaa}\left.+
\frac{(s-p)^2-(s+p)^2\frac{\partial}{\partial x}
\frac{(s-p)^2}{(s+p)^2}}{(s+p)^2 e^x-(s-p)^2}
\right].
\label{26}
\ees

The expansion of the first term under the integral in
powers of the parameter $\alpha$ 
introduced in (\ref{20}) is
\bes
&&
\frac{(s-p\varepsilon)^2-(s+p\varepsilon )^2
\frac{\partial}{\partial x}\frac{(s-
p\varepsilon)^2}{(s+p\varepsilon )^2}}{(s+
p\varepsilon)^2 e^x-(s-p\varepsilon )^2}
=
\frac{1}{e^x-1}\left\{\vphantom{\frac{A}{p^4}}
1+\frac{4}{px}(1-Ax)\alpha \right.
\nn\\
&&\phantom{aaaaa}
+
\frac{8A}{p^2x}(-2-x+2Ax)\alpha^2
+\frac{2}{p^3x}\left[\vphantom{A^2}
2-6p^2+3p^4\right.
\label{27}\\
&&\phantom{aaaaa}\left.
+Ax(-6+32A-32A^2+2p^2-p^4)
+16A(2A-1)\right]\alpha^3
\nn\\
&&\phantom{aaaaa}
+\frac{8A}{p^4x}\left[-8+32A-32A^2+8p^2-4p^4\right.
\nn\\
&&\phantom{aaaaaaaa}\left.\left.+
x(2A-1)(2-16A+16A^2-2p^2+p^4)\right]\alpha^4
+O(\alpha^5)
\vphantom{\frac{A}{p^4}}\right\}.
\nn
\ees 

In the same way for the second term under the integral
of (\ref{26}) one obtains
\bes
&&
\frac{(s-p)^2-(s+p)^2\frac{\partial}{\partial x}
\frac{(s-p)^2}{(s+p)^2}}{(s+p)^2 e^x-(s-p)^2}
=
\frac{1}{e^x-1}\left[
1+\frac{4}{x}(1-Ax)p\alpha \right.
\label{28}\\
&&\phantom{aaaaa}
+
\frac{8A}{x}(-2-x+2Ax)p^2\alpha^2
+\frac{2}{x}\left(\vphantom{A^2}
-1-16A+32A^2-5Ax\right.
\nn\\
&&\phantom{aaaaa}
\left.+32A^2x-32A^3x\right)
p^3\alpha^3
+\frac{8A}{x}\left(-4+32A-32A^2-x\right.
\nn\\
&&\phantom{aaaaaaaa}\left.\left.
+18Ax-48A^2x+32A^3x\right)p^4\alpha^4
+O(\alpha^5)
\right].
\nn
\ees 

Substituting (\ref{27}), (\ref{28}) into (\ref{26})
we firstly calculate integrals with respect to $p$.
All integrals with respect to $x$ are of the form
(\ref{23}). Calculating  them we come to the
following result after long but straightforward 
calculations
\bes
&&
F_l(a)=F_l^{(0)}(a)\left[
1-4\frac{\delta_0}{a}+
\frac{72}{5}\frac{\delta_0^2}{a^2}-
\frac{320}{7}\left(1-\frac{\pi^2}{210}\right)
\frac{\delta_0^3}{a^3}\right.
\nn\\
&&\phantom{aaaaaaaaaaaaaaaa}
\left.+
\frac{400}{3}\left(1-\frac{163\pi^2}{7350}\right)
\frac{\delta_0^4}{a^4}\right],
\label{29}
\eesn
where $F_l^{(0)}(a)\equiv -(\pi^3\hbar cR)/(360a^3)$.
Note that the first order correction from (\ref{29})
was firstly published in [3] and the second order one
 in [17].

Although the results (\ref{24}) and (\ref{29}) for two
configurations were obtained
independently they can be tied by the use of PFT.
By way of example, the energy density
associated with the fourth order contribution 
in (\ref{24}) is
\beq
E_p^{(4)}(a)=
\int\limits_{a}^{\infty}F_p^{(4)}(a)da=
-\frac{5\pi^2\hbar c }{27}
\left(1-\frac{163\pi^2}{7350}\right)
\frac{\delta_0^4}{a^7}.
\label{30}
\eeqn
Then the fourth order contribution to the force between
a plate and a lens 
\beq
F_l^{(4)}(a)=2\pi RE_p^{(4)}(a)=
-\frac{10\pi^3\hbar c R}{27a^3}
\left(1-\frac{163\pi^2}{7350}\right)
\frac{\delta_0^4}{a^4}
\label{31}
\eeqn
 agrees with (\ref{29}). The other coefficients
of (\ref{29}) can be verified in the same way.

\section{Comparision with numerical calculations}

\hspace*{\parindent}
In this section we consider the application range of the
expressions (\ref{24}) and (\ref{29}) for the Casimir force
which take into account higher order conductivity
corrections. It is apparent that the greater the distance
$a$ between the test bodies, the more exact are the
perturbation formulas obtained up to the fourth power in
small parameter $\delta_0/a$. Let us compare the
correction to the Casimir force between two plane parallel
plates given by Eq.~(\ref{24}) with the numerical
results. These results were obtained in \cite{21} for
three metals ($Au,\>Cu$ and $Al$) by the numerical
integration of the formulas which are equivalent to
(\ref{18}). In doing so the tabulated data [28]
for the complex dielectric permittivity was used. The
quantity $\varepsilon(i\xi)$ was obtained through the
imaginary part of dielectric permittivity 
 by the use of dispersion
relation \cite{26}.

In Fig.~1a, the solid line represents computational
results of Ref.~\cite{21} for 
$F_p/F_p^{(0)}$ in
case of $Al$ depending on distance between the plates
$a$. The short-dashed line is obtained from 
Eq.~(\ref{24}) with the value of plasma wavelength
$\lambda_p^{Al}=98\,$nm; the long-dashed line takes
account the terms of (\ref{24}) up to the second power
only. It is seen that (\ref{24}) is in
excellent agreement with computational results of \cite{21}
for all $a\geq\lambda_p^{Al}$. For example, for
$a=0.1\,\mu$m, $0.5\,\mu$m and $3\,\mu$m it follows from
(\ref{24}) that $F_p/F_p^{(0)}=0.56$, 0.85 and 0.97 
which can be compared with computations of
\cite{21}: 0.55, 0.85, and 0.96, respectively.

In Fig.~1b, the analogical results for $Cu$ and $Au$ are
shown. The dashed lines were obtained with
$\lambda_p^{Cu,Au}=132\,$nm. For the typical distances
indicated above it follows from (\ref{24}) that
$F_p/F_p^{(0)}=0.60$, 0.81 and 0.96 which can be
compared with the values: 0.48, 0.81 and 0.96 \cite{21}.
The difference in the first values is due to
$\lambda_p^{Cu,Au}>100\,$nm, i.e. (\ref{24}) is not
applicable for $a=100\,$nm in case of $Cu$ and $Au$.
For $a\geq\lambda_p^{Cu,Au}$
the results agree perfectly well.
 Note that the values
of plasma wavelength 
$\lambda_p=c\sqrt{\pi m}/(e\sqrt{N})$, where $m$ is the
effective mass of conduction electrons, $N$ is their
density are known not very precisely. For $Al$, usually
$\lambda_P^{Al}=100\,$nm is used \cite{28}. For $Au$ and
$Cu$ the value $\lambda_p^{Cu,Au}=136\,$nm was estimated
recently \cite{21}. We used a bit different values which
provide the smallest rms deviation between
the computational results and the ones obtained from
(\ref{24}) (in \cite{28a} $\lambda_p^{Cu}=132\,$nm). 
The values of $F_p/F_p^{(0)}$ at typical
distances do not depend on the change
of $\lambda_p$ for 2--3 percent which is the uncertainty
of the current information regarding $\lambda_p$.

Now let us turn to the Casimir force between a plate and
a lens. The numerical results were obtained in \cite{21}
by the integration of equation equivalent to (\ref{25}).
In~Fig.~2a, the results for $Al$ bodies are shown, and
in Fig.~2b --- for $Cu$ or $Au$ ones. Solid lines
represent computations of \cite{21}, short- and
long-dashed ones are obtained from Eq.~(\ref{29}) used in
full or up to the second power terms.
In both figures the fourth-order perturbation results are
in excellent agreement with computations for all
$a\geq\lambda_p$. At the distances
$a=0.1\,\mu$m, $0.5\,\mu$m and $3\,\mu$m in 
the case of $Al$
we have $F_l/F_l^{(0)}=0.62$, 0.89, 0.98 from
Eq.~(\ref{29}) and 0.63, 0.88, 0.97 from \cite{21}.
For $Cu$ and $Au$ Eq.~(\ref{29}) gives 
 $F_l/F_l^{(0)}=0.59$, 0.85, 0.97 in agreement
with the values: 0.55, 0.85, 0.97 \cite{21}.

As was mentioned in the Introduction the computation of
finite conductivity corrections to the Casimir force by
the use of tabulated data was firstly performed in
\cite{19}. It should be emphasized that our analytical
results are in contradiction with \cite{19}.
By way of example, at $a=0.5\,\mu$m for $Au$ and $Cu$
one can find in \cite{19}
 $F_p/F_p^{(0)}=0.657$ and 0.837 correspondingly 
whereas according to our results  $F_p/F_p^{(0)}=0.81$
for both metals. At the same distance and metals for
 a lens above a plate  
$F_l/F_l^{(0)}=0.719$ and 0.874 \cite{19} whereas from
Eq.~(\ref{29}) one gets 
 $F_l/F_l^{(0)}=0.85$. Our results, however, are in good
agreement with the alternative computations of \cite{21}
supporting the conclusion of \cite{21} that the
manipulation of optical data in \cite{19} is invalid.

It might be well to compare also the exact third and
fourth order conductivity corrections obtained above
with the approximate ones obtained by the use of
interpolation formula \cite{7,10}. To take one example, 
for the force between a lens and a plate the coefficients
near the third and fourth order corrections in
interpolation formula are $-50.67$ and $+177.33$ (compare
with $-43.57$ and $+104.13$ from (\ref{29})). For the
smallest separations $a=120\,$nm in experiment \cite{4}
and $\delta_0/a\approx 0.13$ for $Al$ this leads to 
the 0.5\% difference only in the results obtained by the
interpolation formula \cite{7} and by (\ref{29}).

\section{Conclusions and discussion}

\hspace*{\parindent}
In the above the third and the fourth order corrections to
the Casimir force due to finite conductivity of the 
metal were calculated analytically in configurations of
two plane parallel plates and a spherical lens (or a
sphere) above a plate. The Casimir forces (\ref{24}),
(\ref{29}) are in
excellent agreement with computations of \cite{21} based 
on the tabulated data for the complex dielectric
permittivity for all distances larger than the effective
plasma wavelength of the test body metal. 
What this means is
that the results (\ref{24}), (\ref{29}) can be reliably
used even for the distances $a$ less than the
characteristic absorption wavelength $\lambda_0$ 
 if $\lambda_p<\lambda_0$ (this is a case, e.g.,
for $Au$ and $Cu$, which are characterized by
$\lambda_0\approx 500\,$nm or for Cr with
$\lambda_p\approx 314\,$nm, $\lambda_0\approx 600\,$nm
\cite{29}).

To obtain the higher order conductivity corrections we
have used the plasma model representing dielectric
permittivity by the Eq.~(\ref{16}). This model does not
take into account relaxation processes. However, the
relaxation parameter is much smaller than the plasma
frequency. As was shown in \cite{21}, relaxation could
play some role only for large distances between plates
$a\gg\lambda_p$ and even there the variation of the
corrections to the Casimir force due to it is smaller
than 2\%. If to take into account that for so large
distances the corrections themselves decrease very quickly
it becomes evident that the influence of relaxation can
be neglected.

In conclusion we would like to stress that both the results
(\ref{24}) and (\ref{29}) 
are of the same accuracy in spite of
the fact that the PFT was used in
(\ref{15}) to obtain (\ref{29}). The thing is that this
theorem is equivalent to the addition method of
calculation of the Casimir force which leads to the
error no larger than $10^{-2}$\% for small deviations
from plane parallel geometry \cite{12,30}. What this means
is that for a sphere or spherical lens of large curvature
radius $R\gg a$ the additional error introduced by the
use of PFT is negligible. Therefore it is possible to
conclude that formulas like (\ref{24}),
(\ref{29}) and the analogical expressions for the other
corrections to the Casimir force can be reliably used 
for confronting theory and experiment at a level of
1\% accuracy.

\section*{Acknowledgements}

V.B.B.~wishes to thank Conselho Nacional de Desenvolvimento
Cient\'{\i}fico e Tecnol\'{o}gico (CNPq) for partial
financial support. G.L.K.~and V.M.M.~are grateful to the
Department of Physics of the Federal University of
Para\'{\i}ba (Brazil) for hospitality.

\newpage
\large
\begin{center} {\Large \bf Figure captions} \end{center}
\begin{tabular}{l p{142mm}}
& \\
{\bf Fig.1.} &
 Correction factors to the Casimir force 
$F_p/F_p^{(0)}$ in configuration
of two plane parallel plates for $Al$ (a) and $Cu$ or $Au$
(b) bodies in dependence of distance
measured in $\mu$m. Solid lines
represent the results of computations [21], short-
and long-dashed lines are obtained by the Eq.~(\ref{24})
up to the fourth and the second power respectively.\\
{\bf Fig.2.} &
 Correction factors to the Casimir force 
$F_l/F_l^{(0)}$ in configuration
of a lens (sphere) above a plate 
for $Al$ (a) and $Cu$ or $Au$
(b) bodies in dependence of distance
measured in $\mu$m. Solid lines
represent the results of computations [21], short-
and long-dashed lines are obtained by the Eq.~(\ref{29})
up to the fourth and the second power respectively.
\end{tabular}
\newpage
\begin{figure}[ht]
\renewcommand{\thefigure}{1,\alph{figure}}
\centerline{\epsffile{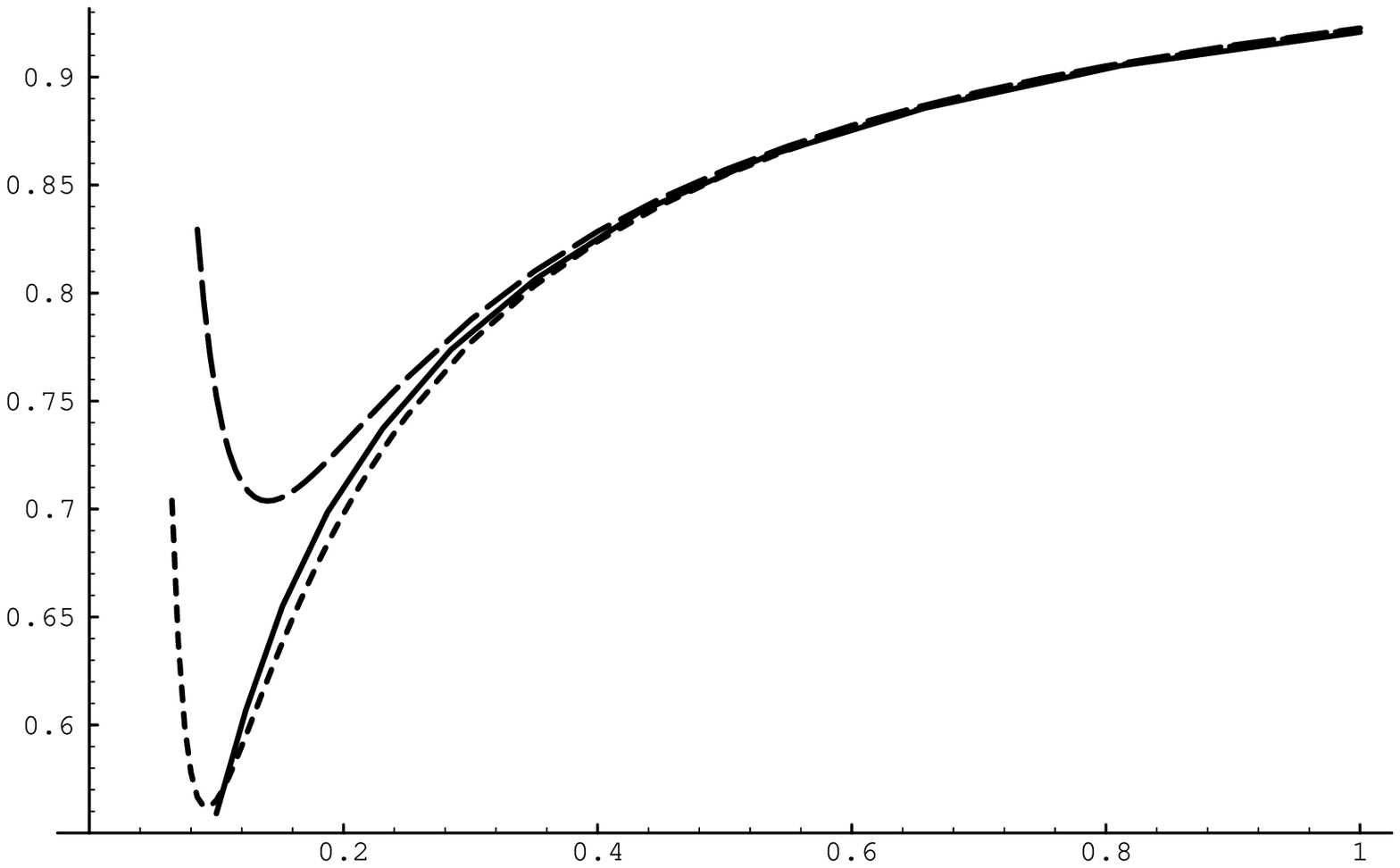}}
\caption{}
\end{figure}
\newpage
\begin{figure}[ht]
\renewcommand{\thefigure}{1,\alph{figure}}
\centerline{\epsffile{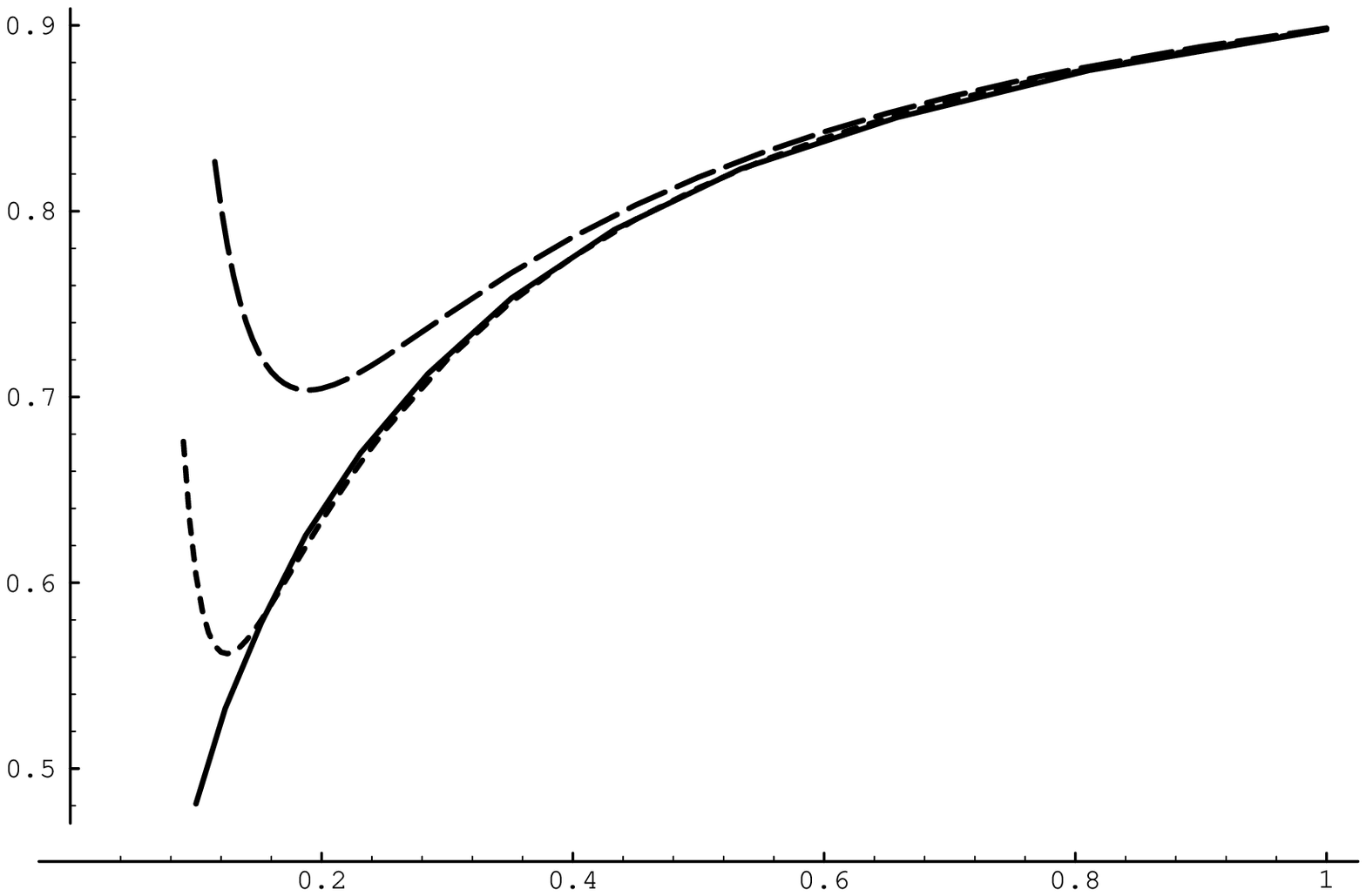}}
\caption{}
\end{figure}
\setcounter{figure}{0}
\newpage
\begin{figure}[ht]
\renewcommand{\thefigure}{2,\alph{figure}}
\centerline{\epsffile{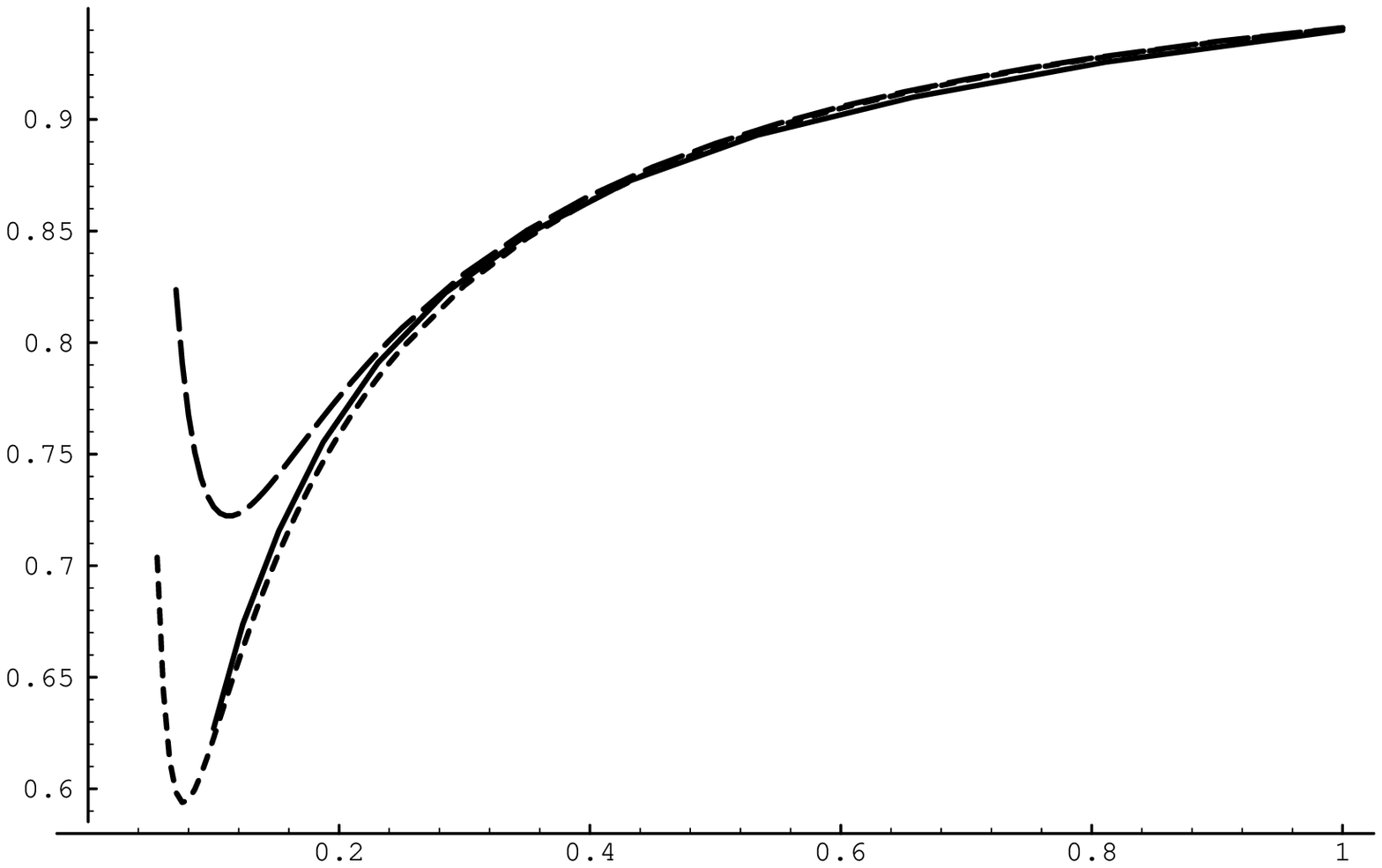}}
\caption{}
\end{figure}
\newpage
\begin{figure}[ht]
\renewcommand{\thefigure}{2,\alph{figure}}
\centerline{\epsffile{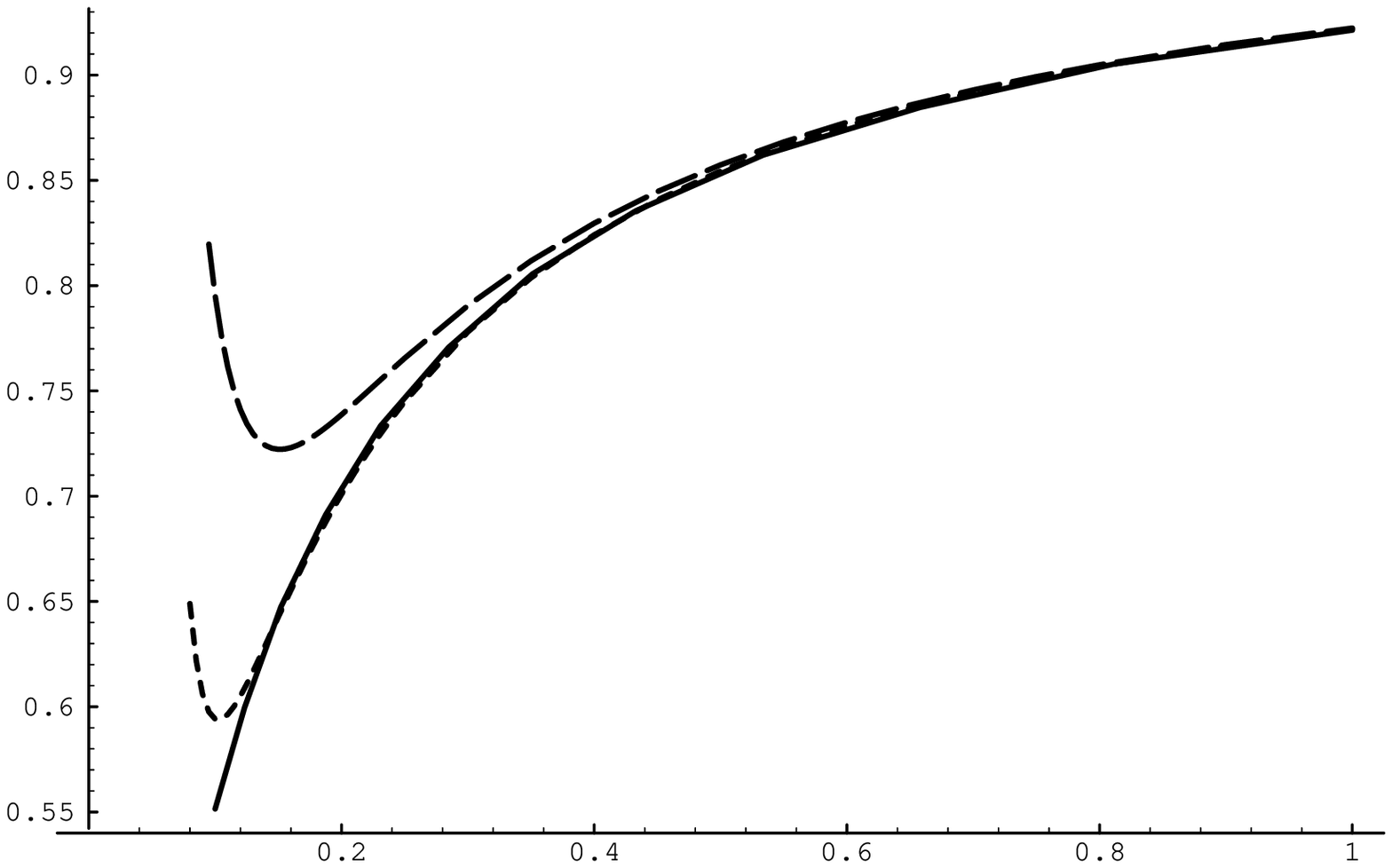}}
\caption{}
\end{figure}

\newpage
\begin{thebibliography}{99}
\bibitem{1}
H.~B.~G.~Casimir, 
 { Proc. Kon. Nederl. Akad. 
Wet.}
{51} (1948) 793.
\bibitem{2}
The Casimir Effect 50 Years Later, ed. M.~Bordag, World
Scientific, Singapore, 1999.
\bibitem{3}
S.~K.~Lamoreaux, Phys. Rev. Lett. 78 (1997) 5;
{81} (1998) 5475.
\bibitem{4}
U.~Mohideen, A.~Roy, 
{ Phys. Rev. Lett.} 
{81} (1998) 4549.
\bibitem{5}
 A.~Roy, U.~Mohideen,
Phys. Rev. Lett. {82} (1999) 4380.
\bibitem{6}
A.~Roy, C.-Y.~Lin, U.~Mohideen, 
Phys. Rev. D 60 (1999) 111101.
\bibitem{7}
G.~L.~Klimchitskaya, A.~Roy, U.~Mohideen,
V.~M.~Mostepanenko, 
Phys. Rev. A 60 (1999) 3487.
\bibitem{8}
G.~L.~Klimchitskaya, E.~R.~Bezerra~de~Mello,
 V.M.~Mostepanenko,
 Phys. Lett. A {236} (1997) 280.
\bibitem{9}
 M.~Bordag, B.~Geyer, G.~L.~Klimchitskaya,
V.~M.~Mostepanenko, 
Phys. Rev. {D 58} (1998) 075003.
\bibitem{10}
 M.~Bordag, B.~Geyer, G.~L.~Klimchitskaya,
V.~M.~Mostepanenko, Phys. Rev. {D 60} (1999)
055004.
\bibitem{11}
J.~C.~Long, H.~W.~Chan, J.~C.~Price, Nucl. Phys.
{ B 539} (1999) 23.
\bibitem{12}
 V.~M.~Mostepanenko, N.~N.~Trunov, 
{The Casimir Effect and its Applications}, 
Clarendon Press, Oxford, 1997.
\bibitem{13}
I.~E.~Dzyaloshinskii, E.~M.~Lifshitz, L.~P.~Pitaevskii,
Sov. Phys. Uspekhi 4 (1961) 153.
\bibitem {14}
C.~M.~Hargreaves, { Proc. Kon. Nederl. Acad. 
Wet.}
{ B68} (1965) 231.
\bibitem {15}
J.~Schwinger, L.~L.~DeRaad,~Jr., K.~A.~Milton,
{ Ann. Phys.} {115} (1978) 1.
\bibitem {16}
V.~M.~Mostepanenko, N.~N.~Trunov, {Sov. J. Nucl.
Phys.} { 42} (1985) 818.
\bibitem{17}
V.~B.~Bezerra, G.~L.~Klimchitskaya, C.~Romero,
Mod. Phys. Lett. A { 12} (1997) 2623.
\bibitem {18}
J.~Blocki, J.~Randrup, W.~J.~Swiatecki, C.~F.~Tsang,
{Ann. Phys.} {105} (1977) 427.
\bibitem{19}
S.~K.~Lamoreaux, Phys. Rev. A 59 (1999) R3149.
\bibitem{20}
S.~K.~Lamoreaux, e-print quant-ph/9907076.
\bibitem{21}
A.~Lambrecht, S.~Reynaud, e-print quant-ph/9907105.
\bibitem{22}
N.~G.~van Kampen, B.~R.~A.~Nijboer, K.~Schram,
Phys. Lett. A 26 (1968) 307.  
\bibitem{23}
P.~W.~Milonni, {The Quantum Vacuum}, Academic Press,
San Diego, 1994.
\bibitem{24}
F.~Zhou, L.~Spruch, Phys. Rev. A 52 (1995) 297.
\bibitem{25}
G.~Barton, Rep. Prog. Phys. 42 (1979) 963.
\bibitem{26}
E.~M.~Lifshitz, L.~P.~Pitaevskii, 
{Statistical Physics, Part 2}, Pergamon Press,
Oxford, 1980.
\bibitem{27}
I.~S.~Gradshteyn, I.~M.~Ryzhik, 
{Table of Integrals, Series and Products},
Academic Press, New York, 1980.
\bibitem{28}
{Handbook of Optical Constants of Solids}, ed.
E.\ D.\ Palik, Academic Press, New York, 1995.
\bibitem{28a}
H.~Ehrenreich, H.~R.~Philipp, Phys. Rev. 128 (1962) 1622.
\bibitem{29}
P.~H.~G.~M.~van Blockland, J.~T.~G.~Overbeek, J. Chem.
Soc. Faraday Trans. 74 (1978) 2637.
\bibitem{30}
 M.~Bordag, G.~L.~Klimchitskaya,
V.~M.~Mostepanenko, 
Int. J. Mod. Phys. {A 10} (1995) 2661. 
\end{thebibliography}
\end{document}